\documentclass[prb,twocolumn,showpacs]{revtex4}

\usepackage{amsmath,bm}
\usepackage{graphicx}

\newcommand{\lr}[1]{\left(  #1 \right)}
\newcommand{\abs}[1]{\left|  #1 \right|}

\newcommand{\Gp}{\Gamma_P}
\newcommand{\Toneeff}{T_1^{\mathrm{eff}}}
\newcommand{\Ttwoeff}{T_2^{\mathrm{eff}}}

\newcommand{\goneeff}{\gamma_1^{\mathrm{eff}}}
\newcommand{\gtwoeff}{\gamma_2^{\mathrm{eff}}}
\newcommand{\goneintr}{\gamma_1^{\mathrm{intr}}}
\newcommand{\gtwointr}{\gamma_2^{\mathrm{intr}}}
\newcommand{\Gptilde}{\widetilde{\Gamma}_P}

\newcommand{\ket}[1]{\left| #1 \right\rangle}

\newcommand{\C}{\mathcal{C}}
\begin{document}

\title{Light narrowing of magnetic resonances in ensembles of nitrogen-vacancy centers in diamond}

\author{K. Jensen, V. M. Acosta, A. Jarmola and D. Budker}
\affiliation{Department of Physics, University of California, Berkeley, California 94720-7300, USA}

\begin{abstract}
We investigate optically detected magnetic resonance signals from an ensemble of nitrogen-vacancy centers in diamond.
The signals are measured for different light powers and microwave  powers, and the contrast and linewidth of the magnetic-resonance signals are extracted.
For a wide range of experimental settings of the microwave and light powers, the linewidth decreases with increasing light power, and more than a factor of two ``light narrowing'' is observed.
Furthermore, we identify that spin-spin interaction between nitrogen-vacancy centers and substitutional nitrogen atoms in the diamond leads to changes in the lineshape and the linewidth of the optically detected magnetic-resonance signals. 
Finally, the importance of the light-narrowing effect for optimizing the sensitivity of magnetic field measurements is discussed.

\end{abstract}
\maketitle

\section{Introduction and motivation}

\subsection{Nitrogen-vacancy centers}
Nitrogen-vacancy (NV) centers in diamond have recently attracted attention within the fields of quantum information science \cite{Wrachtrup06,Dutt07} and optical magnetometry \cite{Taylor08}. Much of the focus has been on single NV centers, and impressive results have been obtained, such as quantum entanglement \cite{Togan10} and nanoscale imaging with single spins \cite{Balasubramanian08,Maze08}. Here we focus on an ensemble of NV centers which can be utilized as a very sensitive $\mu$m to mm sized sensor for magnetic fields \cite{Taylor08, Acosta09PRB, Acosta10APL, Sage12}.

\subsection{Optical magnetometry}
In optical magnetometry \cite{Budker07}, magnetic resonance between two ground states, which are split by a magnetic field $B$, is excited with 
a properly modulated light or magnetic field, where the modulation typically  corresponds to radio-frequency or microwave (MW) range. The magnetic resonance is  measured optically and the magnetic field is determined from the center modulation frequency $\nu_0$ corresponding to the magnetic resonance.
To be able to determine the center frequency 
of the optically detected magnetic resonance (ODMR) signal 
and thereby the magnetic field with high precision, a narrow linewidth $\Delta \nu$ is desired. 
The magnetometer signal is proportional to $\C/\Delta \nu$, where $\C$ is the contrast of the ODMR signal. 
If the main source of noise is shot noise of the detected light, the magnetic field-sensitivity (measured in $\mathrm{T}/\sqrt{\mathrm{Hz}}$) and also the noise-to-signal ratio will be proportional to $\Delta \nu/\lr{\C \sqrt{P}}$, where $P$ is the light power. 
Both the contrast and the linewidth depend on the MW power and the light power. 
To achieve high magnetic field-sensitivity one therefore needs to optimize the MW power and the light power.

\subsection{Light narrowing}
The dependence of the magnetometer signal on MW power and light power is now discussed.
In general, both the contrast and the linewidth increase with MW power.
In atom-light interactions, typically, the linewidth also increases with increasing light power. 
However, in some situations, the linewidth 
can decrease with increasing light power. 
This phenomenon is called light narrowing, with an early observation 
in an optically pumped 
alkali vapor
reported in Ref.\ \onlinecite{Bhaskar81}.
In that work power broadening due to the MW field was reduced when the light power was increased. 
A similar
light-narrowing effect on NV centers in diamond is the topic of this paper.
A different type of light narrowing, which occurs even in the absence of MW power broadening, was demonstrated in Ref.\ \onlinecite{Appelt99} and \onlinecite{Scholtes11}.
There the linewidth was reduced due to the suppression of spin-exchange collisions when the alkali vapor became more and more polarized with increasing light power. 
This latter effect, however, does not appear to have a direct relation to this work.

\section{Theory}

\subsection{The Bloch formula}
Light narrowing of a power-broadened magnetic resonance can be understood using a two-level model pictured in Fig.\ \ref{fig:setuplevels}(a). Consider a single NV center (or atom) with two relevant ground states $\ket{0}$ and $\ket{1}$ which are coupled by a MW magnetic field with coupling strength given by the on-resonance Rabi frequency $\Omega_R\propto \sqrt{P_{\rm{MW}}}$, where $\Omega_R$ is measured in rad/s and $P_{\rm{MW}}$ is the MW power.
The longitudinal and transverse relaxation times of the atom are denoted $T_1$ and $T_2$, and the corresponding decay rates are denoted $\gamma_1=1/T_1$ and $\gamma_2=1/T_2$.
We assume that the system can be optically pumped into the state $\ket{0}$ with a ``pump rate''  $\Gp$.
For a continuous-wave MW magnetic field with detuning $\Delta$ (in rad/s), one calculates  using the Bloch equations (Appendix A) that the magnetic-resonance signal will 
be Lorentzian as a function of MW detuning [Eq.\ (\ref{eq:2levellineshape})]. After converting from rad/s to Hz, we find the signal as a function of frequency $\nu$ in Hz to be
\begin{equation}
S(\nu-\nu_0)=S(\infty)
\left[ 1-\frac{\C\cdot \lr{\Delta \nu/2}^2}{\lr{\nu-\nu_0}^2+ \lr{\Delta \nu/2}^2} \right].
\label{eq:SDelta}
\end{equation}
$S(\infty)$ is the signal when the MW's are off-resonant 
($\abs{\nu-\nu_0} \rightarrow \infty$), $\C$ is the contrast of the resonance, and $\Delta \nu$ is the full width at half maximum (FWHM) measured in Hz. 
In our experiment, the signal is the detected fluorescence from the NV centers.
The signal is maximal off-resonance because optical pumping of the NV center leads to increased population of the $\ket{0}$ state which fluoresces stronger than the $\ket{1}$ state. When the MW field is on-resonance, population is transfered from the $\ket{0}$ state to the $\ket{1}$ state leading to reduced fluorescence.

The width of the resonance $\Delta \nu$ can be expressed as 
\begin{equation}
\Delta \nu=\sqrt{\lr{\frac{1}{\pi\Ttwoeff}}^2+\lr{\frac{4\Toneeff}{\Ttwoeff}}\cdot f_R^2},
\label{eq:Blochformula}
\end{equation}
where $f_R=\Omega_R/\lr{2\pi}$ is the Rabi frequency measured in Hz.
The ``effective'' relaxation times depend on the optical-pumping rate: $1/\Toneeff=1/T_1+\Gp$ and $1/\Ttwoeff=1/T_2+\Gp/2$.
For low MW power 
the width of the magnetic resonance is approximately 
$\Delta \nu \approx 1/\lr{\pi \Ttwoeff}$. Increasing the MW power leads to power broadening, and for high MW power the FWHM 
\begin{equation}
\Delta \nu \approx 2 \cdot \sqrt{\Toneeff/\Ttwoeff} \cdot f_R .
\end{equation}
equals the Rabi frequency  magnified by the factor 
$2\cdot \sqrt{\Toneeff/\Ttwoeff}$.

For an ensemble of NV centers in type 1b diamond (nitrogen density $\approx 100$ ppm), representative values for room-temperature intrinsic relaxation times 
(i.e., in the absence of light and microwaves) 
are $T_1^{\rm{intr}}\approx 1 \;\rm{ms} \;$  \cite{Jarmola2012} and $T_2^{\rm{intr}} \approx 1 \; \mu \rm{s} \;$  \cite{Pham2012} and thereby differ by 3 orders of magnitude.
In our experiment, laser light is used to probe the NV centers and simultaneously optically pump the NV centers into the state $\ket{0}$. 
This optical pumping reduces $\Toneeff$.
In general, laser light also leads to a reduced $\Ttwoeff$, but for certain ``low'' light powers, we can be in a regime where $\Ttwoeff$ is almost unaffected by the light but $\Toneeff$ is significantly reduced. For a power broadened magnetic-resonance line, the width will therefore decrease with increasing light power since the width is proportional to 
$\sqrt{\Toneeff/\Ttwoeff}$.

\subsection{Inhomogeneous broadening}
Equation (\ref{eq:Blochformula}) is valid for a single two-level atom or for an ensemble of identical two-level atoms. However, the NV centers in the ensemble are different since they are situated in different local environments. This inhomogeneous broadening can be modelled as though the NV $\ket{0} \leftrightarrow \ket{1}$ resonance frequency $\nu_0$ has a probability distribution $P\lr{\nu_0}$ 
[the probability per unit frequency for the resonance frequency $\nu_0$ to be at this value]
with FWHM $\Delta \nu_{\rm{inh}} \propto 1/ T_2^{*}$.
In general  $T_2 \leq 2T_1$ and typically $T_2^{\rm{intr}} \gg T_2^*$ in solids.
The ensemble averaged ODMR signal is then
\begin{equation}
S_{\rm{tot}}(\nu)=\int_0^{\infty} P(\nu_0)  S(\nu-\nu_0) d\nu_0 ,
\label{eq:Sintegrated}
\end{equation}
where $S(\nu-\nu_0)$ is the ODMR signal [Eq.\ (\ref{eq:SDelta})] for a single NV center with resonance frequency $\nu_0$.
The FWHM $\Delta \nu_{\mathrm{tot}}$ of the ensemble averaged ODMR signal can be found by calculating the FWHM of $S_{\rm{tot}}(\nu)$ in Eq.\ (\ref{eq:Sintegrated}).
If the resonance frequency has a Gaussian probability distribution:
$P(\nu_0)\propto \exp\lr{-\nu_0^2/2\sigma^2}$ 
[where $\sigma$ is defined such that the FWHM
$\Delta \nu_{\rm{inh}}\approx 2.35 \sigma$],
the lineshape of $S_{\rm{tot}}(\nu)$ will be the Voigt lineshape \cite{Lasers} which can be calculated using complex error functions.
If on the other hand, the resonance frequency has a Lorentzian probability distribution, the lineshape of $S_{\rm{tot}}(\nu)$ will also be Lorentzian with a FWHM given by 
\begin{equation}
\Delta \nu_{\rm{tot}}=\Delta \nu_{\rm{inh}}+\Delta \nu,
\label{eq:sumofwidths}
\end{equation} 
where $\Delta \nu_{\rm{inh}}=1/\lr{\pi T_2^{*}}$ and $\Delta \nu$ is given by Eq.\ (\ref{eq:Blochformula}).
Although it is probably most natural to assume that $P(\nu_0)$ is Gaussian, we will assume that $P(\nu_0)$ is Lorentzian.
This is done mainly because one can obtain a simple analytical formula for the total width 
[Eq.\ (\ref{eq:sumofwidths})], which later can be used for fitting the experimental data. Also, it seems that the measured ODMR spectra can be fitted well with Lorentzian functions (see for instance Fig.\ \ref{fig:ODMR3}b).

Inserting the Bloch formula given by Eq.\ (\ref{eq:Blochformula}) into Eq.\ (\ref{eq:sumofwidths}) we find
\begin{eqnarray}
\Delta \nu_{\rm{tot}} & \approx & \Delta \nu_{\rm{inh}}+
\sqrt{\lr{\frac{\gtwointr}{\pi}}^2+\lr{\frac{4\gtwointr}{\gamma_1+\Gamma_P}}\cdot f_R^2} \nonumber \\
&\approx & \Delta \nu_{\rm{inh}}+ f_R \cdot
\sqrt{\frac{4\gtwointr}{\gamma_1+\Gamma_P}} ,
\label{eq:inhwidth}
\end{eqnarray} 
where we first assumed that the light does not affect the transverse relaxation rate $\gtwoeff \approx \gtwointr$, and next omitted the term $\gtwointr/\pi$ because it is much smaller than $\Delta \nu_{\rm{inh}}$ and will be hard to estimate from the width of the inhomogeneously broadened ODMR lineshape.

\section{Experiment}

\subsection{NV center level structure} 
NV centers in diamond occur in at least two charge states, NV$^0$ and NV$^-$, where the negatively charged center is of primary interest since it is possible to optically polarize and readout the NV$^-$ spin state.
Figure \ref{fig:setuplevels}(b) shows the level structure of the 
NV$^-$ center.
The ground state is a spin-triplet $^3A_2$ and the states $\ket{m_s=0}$ and $\ket{m_s=\pm1}$ (the quantum number $m_s$ is the electron-spin projection along the NV axis) have a zero-magnetic-field splitting of 2.87 GHz. Applying a static magnetic field along the NV axis leads to a splitting of the $\ket{m_s=+1}$ and $\ket{m_s=-1}$ states. 
The ground states are coupled to a spin-triplet excited state $^3E$ using laser light. The zero-phonon line (ZPL) occurs at 637 nm, however, typically the NV center is excited with green light (532 nm) through the phonon sideband (PSB).
Once excited, an NV$^-$ center can decay back to the ground states directly (through the spin-conserving transition $^3A_2 \leftrightarrow$ $^3E$) emitting red fluorescence in the PSB (with wavelength $\approx$ 637-800 nm) or through a spin non-conserving channel through the spin-singlet states $^1A_1$ and $^1E \;$ \cite{Rogers2008,Acosta10PRB}.
(The order of the singlet levels shown in Figure \ref{fig:setuplevels}(b) is given in accordance with the recent work \cite{Doherty11,Choi2012}.)
The latter channel through the singlet levels is more probable for the $\ket{m_s=\pm1}$ states than for the $\ket{m_s=0}$ state of $^3E$.
This leads to the possibility to efficiently optically pump, polarize and read-out the NV$^-$ spin state.


\subsection{Setup and procedure}

\begin{figure}
	\centering
\includegraphics[width=.47\textwidth]{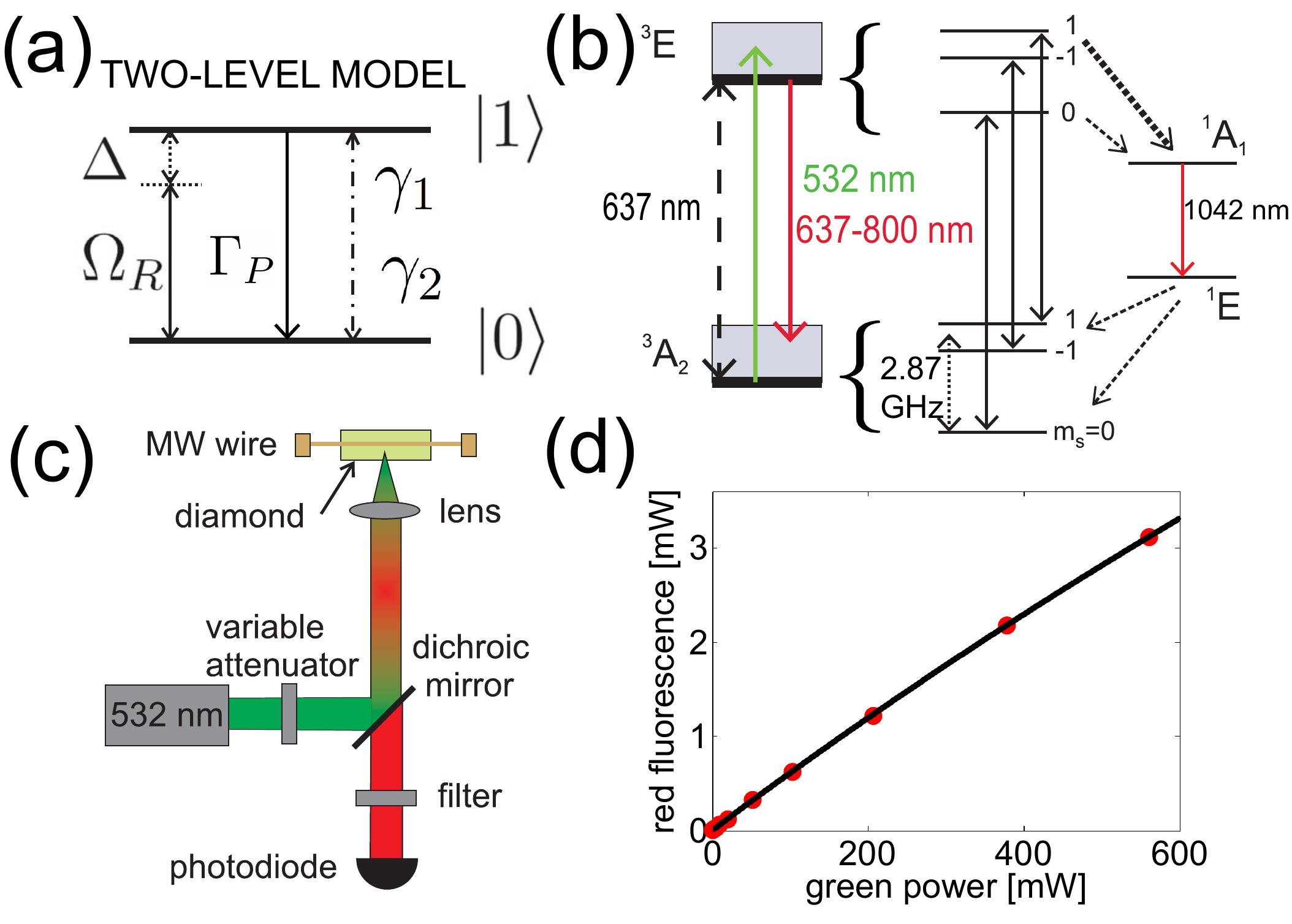}
\caption{
(a) Two-level model of ODMR dynamics. Two states $\ket{0}$ and $\ket{1}$ are coupled with a MW field (Rabi frequency $\Omega_R$ and detuning $\Delta$); $\Gamma_P$  optical-pumping rate; $\gamma_1$ longitudinal spin-relaxation rate;
$\gamma_2$ transverse spin-relaxation rate. 
(b) Level scheme of the NV$^-$ center.  
(c) Schematic of the confocal setup.
(d) Points: Measured red fluorescence $P_{fl}$ as a function of green pump power $P$. Solid line is a fit to a function of the form
$P_{fl} = k \cdot P/(1+P/P_{sat})$.}
\label{fig:setuplevels}
\end{figure}

The setup is shown in Fig.\ \ref{fig:setuplevels}(c). 
We use the high pressure high temperature HPHT diamond sample S5 (investigated previously in Ref.\ \onlinecite{Acosta09PRB}) with the following properties: nitrogen concentration $\approx 200$ ppm, NV$^-$ concentration $\approx$ 12 ppm, and a cut perpendicular to the [100] crystallographic direction.
The diamond is located in a static magnetic field produced 
by a permanent magnet.
A MW generator sending current through a wire located on top of the diamond produces the MW field. 
Green (532 nm) light from a laser
is focused onto the diamond using a confocal setup with an aspheric lens with focal length of 4 mm and numerical aperture NA=0.60, and the red flouresence is detected with a photodiode after passing through a dichroic mirror.
The amount of red fluorescence as a function of green light power is plotted in 
Fig.\ \ref{fig:setuplevels}(d) together with a fit of the form 
$P_{fl} = k \cdot P/(1+P/P_{sat})$ where $k=6.21(2) \cdot 10^{-3}$ and $P_{sat}=4.8(2)$ W are fitted parameters. 
The saturation power for equalizing the populations of the ground ($^3A_2$) and excited ($^3E$) states is denoted $P_{sat}$.
For the light powers investigated here ($P \leq 500$ mW), we hardly see any saturation. 
The MW frequency is scanned and the ODMR signal is recorded using an oscilloscope. 
Four different orientations of the NV axis in the diamond lattice are possible leading to eight magnetic resonances (since each of the four orientations has two electron-spin resonances: $\ket{m_s=0} \leftrightarrow \ket{m_s=+1}$ and $\ket{m_s=0} \leftrightarrow \ket{m_s=-1}$).
A static magnetic field of 77 G is 
aligned along one of the $[111]$ crystallographic directions. In this case, NV centers oriented along this axis have magnetic-resonance frequencies which are different from the overlapping resonances of the three other orientations. For this magnetic-field direction, one should therefore observe four magnetic-resonance frequencies.

\subsection{Optically detected magnetic resonance signals}

\begin{figure*}
	\centering
\includegraphics[width=\textwidth]{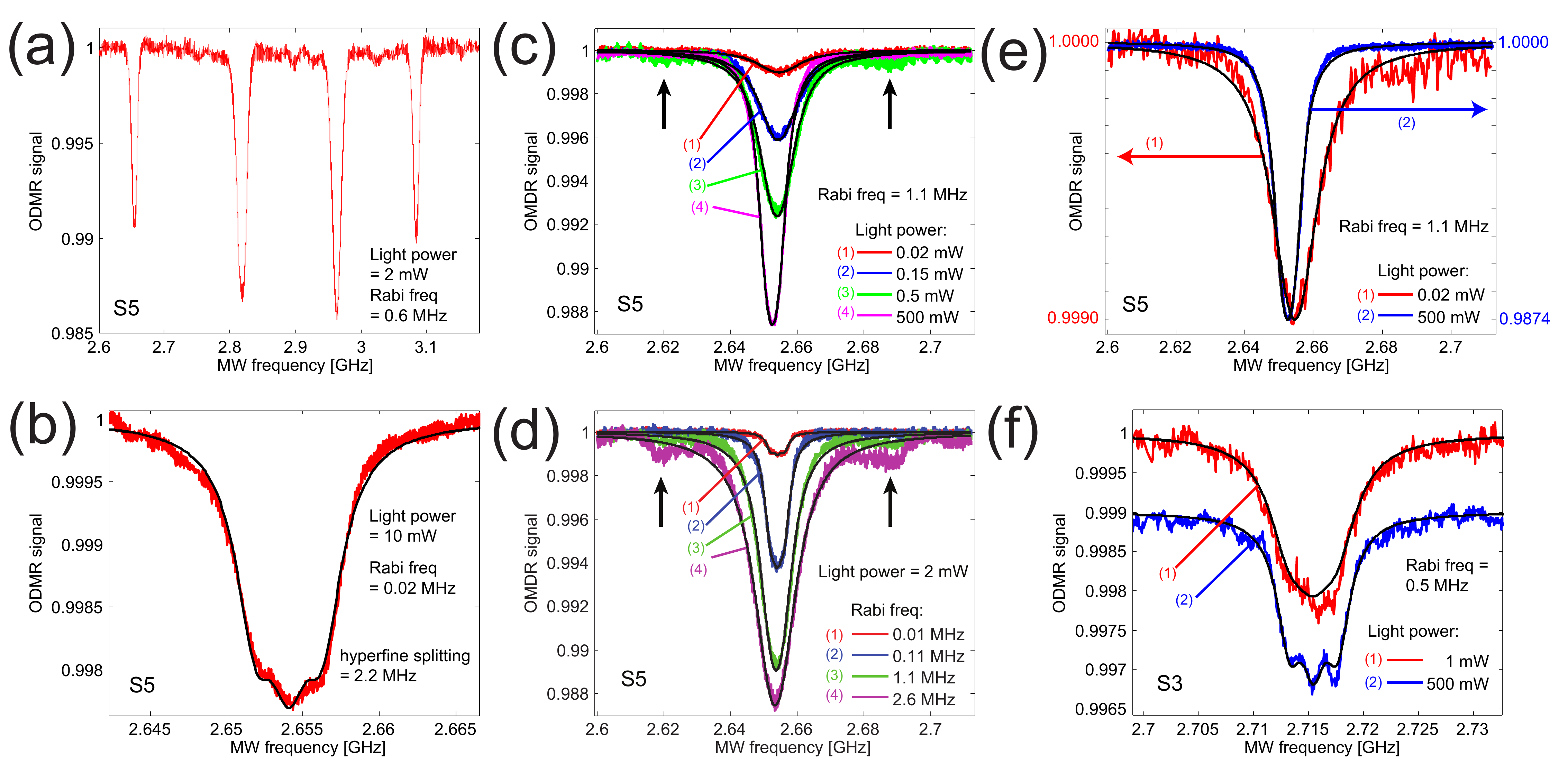}
\caption{
Examples of ODMR signals taken with the S5 (a)-(e) and S3 (f) diamond samples at rooom temperature. 
(a) ODMR signal with a wide scan range of the MW frequency. 
(b) Narrow scan range ODMR signal of the magnetic resonance centered around 2654 MHz.
(c) Light power fixed and four different settings of the MW power. 
(d) MW power fixed at a relatively high value and four different settings of the light power.
(e) Demonstration of light narrowing with the S5 sample. Notice the two y-axes.
(f) Demonstration of light narrowing with the S3 sample. The bottom spectrum ($P=500$ mW setting) has been displaced vertically by -0.001.}
\label{fig:ODMR3}
\end{figure*}

ODMR signals were recorded for different settings of the MW and light powers. 
To determine the strength of the MW field, we employed a pulsed calibration measurement where the on-resonance Rabi frequency was measured for different MW settings. We found the expected $f_R \propto \sqrt{P_{MW}}$ square-root dependence of the Rabi frequency on the MW power.

Examples of ODMR signals are shown in Fig.\ \ref{fig:ODMR3}. The ODMR signals are normalized to the signal obtained when the MW frequency is far detuned. 
Figure \ref{fig:ODMR3}(a) shows an ODMR signal for a wide scan range of the MW frequency. Four magnetic-resonance frequencies of the NV center are observed, as expected when the magnetic field is along one of the [111] directions. 
We now focus on the magnetic resonance with the lowest resonance frequency,
which corresponds to the $\ket{m_s=0} \leftrightarrow \ket{m_s=-1}$ transition of NV centers oriented along the magnetic field direction.
A narrower scan of the lowest resonance is shown in Fig.\ \ref{fig:ODMR3}(b). The MW power was set at a low value for this measurement (corresponding to $f_R=0.02$ MHz).
In this case, the resonance is sufficiently narrow to partially resolve the hyperfine structure of the NV center.
[The hyperfine interaction of the NV electron spin and the $^{14}N$ nuclear spin $I=1$ splits each 
$\ket{m_s=0} \leftrightarrow \ket{m_s=\pm1}$
resonance into three resonances with splitting $A_{\rm{hf}}=2.2$ MHz (see for instance Fig.\ 10 in Ref.\ \onlinecite{Acosta09PRB}).]

Figure \ref{fig:ODMR3}(c) shows ODMR signals for different settings of the MW power. 
The central resonance frequency $\nu_0\approx 2654$ MHz shifts slightly to lower frequencies with increasing MW power due to heating \cite{Acosta2010}. 
Also, one clearly sees that increasing the MW power leads to power broadening of the ODMR signal.

We now proceed with the demonstration of light narrowing. 
Figure \ref{fig:ODMR3}(d) shows examples of ODMR signals for four different settings of the light power $P$ ranging from 0.02 to 500 mW. The MW Rabi frequency was at a relatively high value of 1.1 MHz. 
Fitting the spectra reveals that the linewidth decreases with increasing light power (the details of the fit procedure and the fit results are discussed later).
This is the light-narrowing effect.
The spectra with the lowest and highest light-power settings are also plotted in 
Fig.\ \ref{fig:ODMR3}(e) [notice the two y-axes in Fig.\ \ref{fig:ODMR3}(e)]. There we clearly see a reduction of the linewidth by approximately a factor of two when the light power is increased from 0.02 mW to 500 mW.

We also demonstrate light narrowing with the S3 diamond sample which has a lower concentration of both $\rm{NV}^-$ centers ($\approx 0.012$ ppm) and nitrogen impurities ($\leq 1$ ppm) (for details of the S3 sample see Ref.\ \onlinecite{Acosta09PRB}). 
Due to the lower impurity concentration, the S3 sample has less inhomogeneous broadening compared to the S5 sample ($T_2^* \approx 300$ ns for the S3 sample compared to $T_2^* \approx 100$ ns for the S5 sample). This makes the hyperfine structure better resolved in the S3 sample compared to the S5 sample.
Figure \ref{fig:ODMR3}(f) shows ODMR spectra for the S3 sample for two settings of the light power.
Notice that the bottom spectrum ($P=500$ mW setting) has been displaced vertically  by -0.001 for better visualization. At the low light power setting, the spectrum is relatively broad and a single resonance is observed. At the high light power setting, the spectrum is narrower, and the hyperfine structure can be partially resolved. 
I.e., spectral features which are not resolved at low light power can be resolved at higher light power due to the light-narrowing effect.

\subsection{Side resonances}
A pair of side resonances  at a distance of $\delta \approx 33$ MHz from the central resonance can be seen in Fig.\ \ref{fig:ODMR3}(c) (marked by arrows). The contrasts of the side resonances increase with increasing Rabi frequency.
The side resonances are due to MW-induced simultaneous spin-flips of NV centers and nearby substitutional nitrogen atoms (also called P1 centers) \cite{vanOort90}. 

Let $\nu_0$ be a MW spin-transition frequency of the NV center  and $\delta$ a transition frequency between two spin states of a nearby P1 center (which has electron spin 1/2 and nuclear spin 1). 
Due to magnetic dipole-dipole interaction between the NV center and the nearby P1 center, the spin eigenstates of the NV center and the P1 center will be slightly mixed.
In this case, it is possible to drive MW transitions with resonance frequencies $\nu_0\pm\delta$ which flip the spin of the NV center and the P1 center simultaneously.

The side resonances were studied in detail and the results will be reported elsewhere \cite{Simanovskaia2012}. In short, many side resonances were observed, with resonance frequencies depending on the magnetic-field amplitude and orientation \cite{vanOort90}. 
The contrasts of the side resonances were found to be sample dependent and to increase linearly with  MW power at low MW powers and saturate at higher MW powers \cite{Simanovskaia2012}.
In the S5 sample [see Fig.\ \ref{fig:ODMR3}(c)], the side resonances were rather small and did not overlap much with the resonance centered at $\nu_0$, but in other samples the side resonances were larger and in some cases they significantly distorted the spectrum of the NV center \cite{Simanovskaia2012}.
We note that  while side resonances were observed in the S5 sample,  no side resonances were observed in the S3 sample. This makes sense since the concentration of P1 centers in the S3 sample ($\leq$ 1 ppm) is much lower than in the S5 sample ($\approx$ 200 ppm) \cite{Acosta09PRB}. We would therefore not expect any visible side resonances in the S3 sample.

The rate of MW-induced simultaneous spin-flips $\gamma_{\mathrm{NV-P1}}$ should depend on several parameters including P1 concentration, MW detuning from the side resonances at $\nu_0\pm \delta$, MW power and light power.
We note that the mechanism behind the MW-induced simultaneous spin-flips is rather complicated. A detailed analysis of how NV-P1 interactions change the ODMR spectrum is beyond the scope of this paper.
We would instead like a simple model for how the MW-induced simultaneous spin-flips of NV and P1 centers influence the width of the magnetic resonance centered at $\nu_0$. 
Spin-flips of the NV electron spin due to interaction with P1 centers can be thought of as a source of $T_1$-relaxation for the NV centers. 
We will assume that the only effect of NV-P1 interaction is to increase the $T_1$-relaxation rate by an amount corresponding to the spin-flip rate: $\gamma_1=\goneintr + \gamma_{\rm{NV-P1}}$.

For simplicity, we will neglect the detuning dependence of the spin-flip rate. This assumption should be valid since the side resonances are well separated from the resonance centered at $\nu_0$, and  since we only are interested in the effect of the simultaneous spin-flips on the width of the resonance centered at $\nu_0$.

Since the contrast of the side resonances increase with increasing MW power, the spin-flip rate should also increase with increasing MW power. The exact dependence on the MW power is not known, but we will  assume that the rate is proportional to  MW power, or equivalent, the square of the Rabi frequency. We will also assume that the rate of spin-flips saturate for high MW powers. 
We end up with the following empirical formula for the spin-flip rate:
\begin{equation}
\gamma_{\mathrm{NV-P1}}=a(P) \cdot f_R^2/\lr{1+f_R^2/f_0^2}
\label{eq:gNVP1}
\end{equation}
The saturation Rabi frequency $f_0$ and the proportionality constant $a(P)$ and its dependence on light power  will be determined from the experimental data.

\subsection{Fitting procedure}
The normalized ODMR signals $S(\nu)$ are fitted to the following function
\begin{equation}
S(\nu)=1-\sum_{m_I=-1}^{1} \frac{A g^2 }{\lr{\nu-\nu_0-m_I \cdot A_{\rm{hf}}}^2+g^2},
\label{eq:ODMRsignal}
\end{equation}
which is a sum of three Lorentzians each with peak height (or amplitude) $A$, center frequency $\nu_0+m_I \cdot A_{\rm{hf}}$ ($m_I=-1,0,1$) and FWHM $2g$. 
This fit function with three components separated by $A_{\rm{hf}}=2.2$ MHz is chosen due to the hyperfine structure of the NV center.

The fits are shown on top of the data in Fig.\ \ref{fig:ODMR3} as solid lines.  From the fits we extract the FWHM $2g$ and the amplitude $A$ of the ODMR signals. Notice that we choose to define the width of the ODMR signal as the width of an individual hyperfine component (which is smaller than the width of the total ODMR signal).
When fitting, we excluded the part of the spectrum where the side resonances are located ($\approx$ 20 MHz of MW scan range centered at each side resonance was excluded). Fitting the whole ODMR spectrum did not change the fitted parameters significantly.

\subsection{Linewidth}

\begin{figure*}[ht]
\centering
\includegraphics[width=0.8\textwidth]{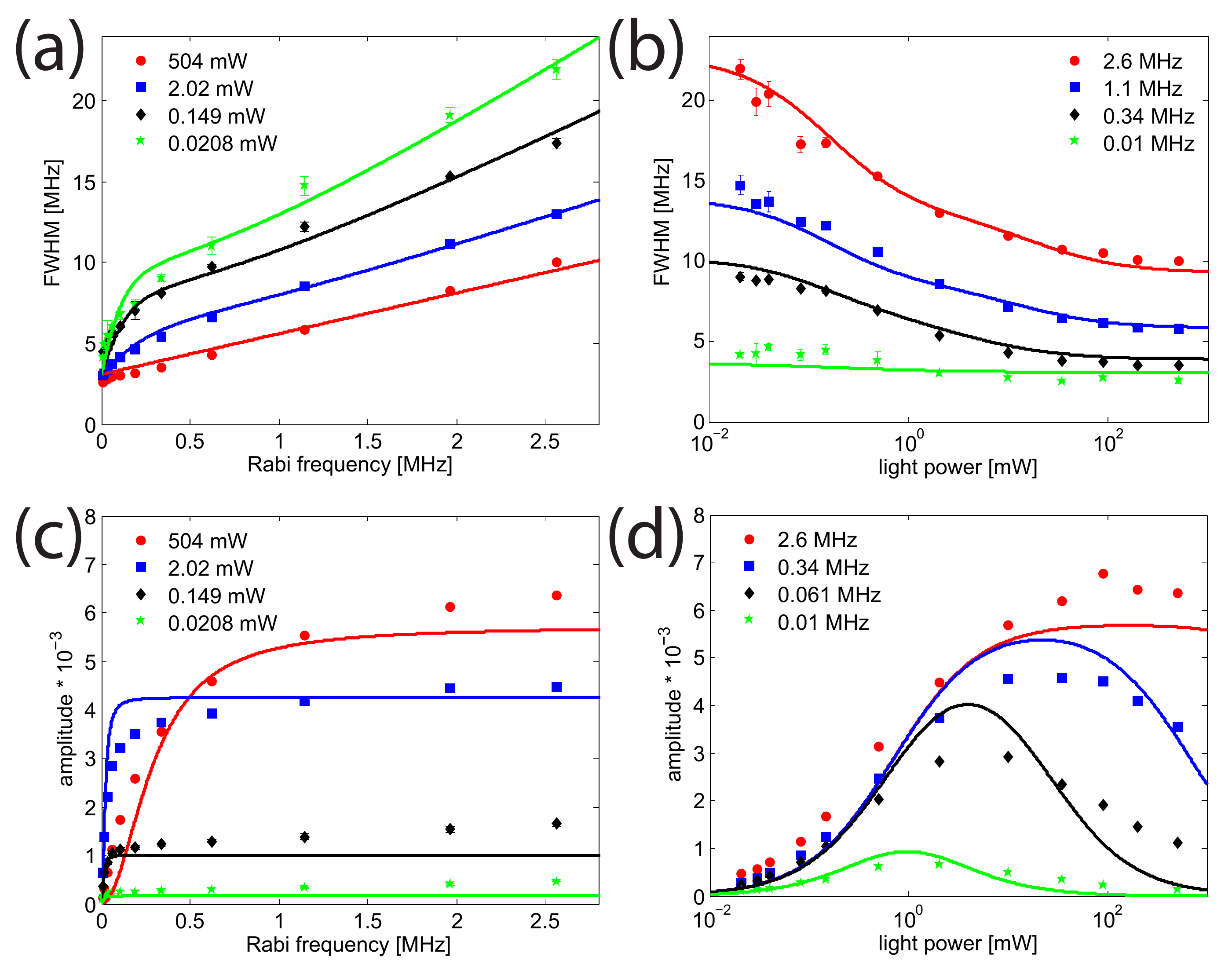}
\caption{Points: Experimental results of the FWHM and amplitude of ODMR signals as a function of Rabi frequency and light power. Solid lines: Global fits to the FWHM and the amplitude. 
The uncertainties on the experimental data points correspond to the standard deviations found from repeated measurements of the width and contrast.}
\label{fig:ResultsAndTheory2012}
\end{figure*}

More ODMR signals were recorded, and the fitted FWHM and amplitude are plotted in 
Fig.\ \ref{fig:ResultsAndTheory2012} as a function of light power and Rabi frequency, respectively. 
The light power was varied more than four orders of magnitude and
we  therefore use logaritmic scale for the power axis.

We will first focus on the dependence of the linewidth on the light power.
Figure \ref{fig:ResultsAndTheory2012}(b) shows the width as a function of the light power for several settings of the Rabi frequency. We observe a significant light narrowing (decreasing width with increasing light power) for all settings of the Rabi frequency. A maximum of 2.5 reduction of the width was found for the higher Rabi frequencies.
Note that in a related study \cite{Dreau11}, where the widths of ODMR signals from single NV's were examined, no light narrowing was observed. 
In that study \cite{Dreau11} the light intensity was high and lead to significant power broadening instead of light narrowing.

For the highest light power ($P=500$ mW), the width seems to have a linear dependence on the Rabi frequency [Fig.\ \ref{fig:ResultsAndTheory2012}(a)].
For lower light powers, we observe a non-linear scaling with Rabi frequency, in particular we found that the width  as a function of Rabi frequency 
has a negative curvature (i.e., the second derivative of the width with respect to Rabi frequency is negative).
According to Eq.\ (\ref{eq:inhwidth}) for the linewidth, the curvature is negative if $\gamma_1$ increases with Rabi frequency. 
We believe that the negative curvature is a consequence of MW-induced simultaneous spin-flips of NV and P1 centers which effectively increases $\gamma_1$ with increasing Rabi frequency.

As a remark, we note that the presence of side resonances was correlated with a negative curvature. In the S5 sample (which has a high density of P1 centers), both side resonances and a negative curvature were observed, while in the S3 sample (which has a low density of P1 centers), neither side resonances nor negative curvature were observed. This supports our hypothesis, that the negative curvature is due to NV-P1 simultaneous spin-flips.

Assuming that the longitudinal spin-relaxation rate is
$\gamma_1 =\goneintr + \gamma_{\mathrm{NV-P1}}$, where $\gamma_{\mathrm{NV-P1}}$ is given by Eq.\ (\ref{eq:gNVP1}), the linewidth of the ODMR resonance 
[Eq.\ (\ref{eq:inhwidth})] is expected to be on the following form:
\small
\begin{equation}
\Delta \nu_{\rm{tot}} \approx \Delta \nu_{\rm{inh}} + f_R \cdot
\sqrt{  
\frac{4\gtwointr \cdot \lr{1+P/P_0}}{\goneintr + 
a(P) \cdot f_R / \lr{1+f_R^2/f_0^2}+c \cdot P}
}. 
\label{eq:widththeory1}
\end{equation}
\normalsize
For low light powers, the optical pumping rate is assumed to be proportional to light power: $\Gamma_P=cP$ with proportionality constant $c$. 
The pump rate should saturate at higher light powers due to the finite lifetime of the singlet states responsible for the optical-pumping mechanism of the NV center. 
The effect of saturation on the width of the ODMR spectrum is analyzed in Appendix B using a five-level model for ODMR dynamics. 
We find that one should include the term $\lr{1+P/P_0}$ in Eq.\ (\ref{eq:widththeory1}) for the width [see Eq.\ (\ref{eq:g5levelPower}) in Appendix B]. 
Note that the optical pumping saturation power $P_0$ which is related to the singlet state lifetime ($\approx 200$ ns at room temperature \cite{Acosta10PRB}) should be smaller than the saturation power $P_{sat}$ [shown in Fig.\ \ref{fig:setuplevels}(d)] for equalizing the populations in the ground and excited states ($P_{sat}$ is related  to the excited state lifetime $\approx 12$ ns).

\begin{figure}[ht]
\centering
\includegraphics[width=0.44\textwidth]{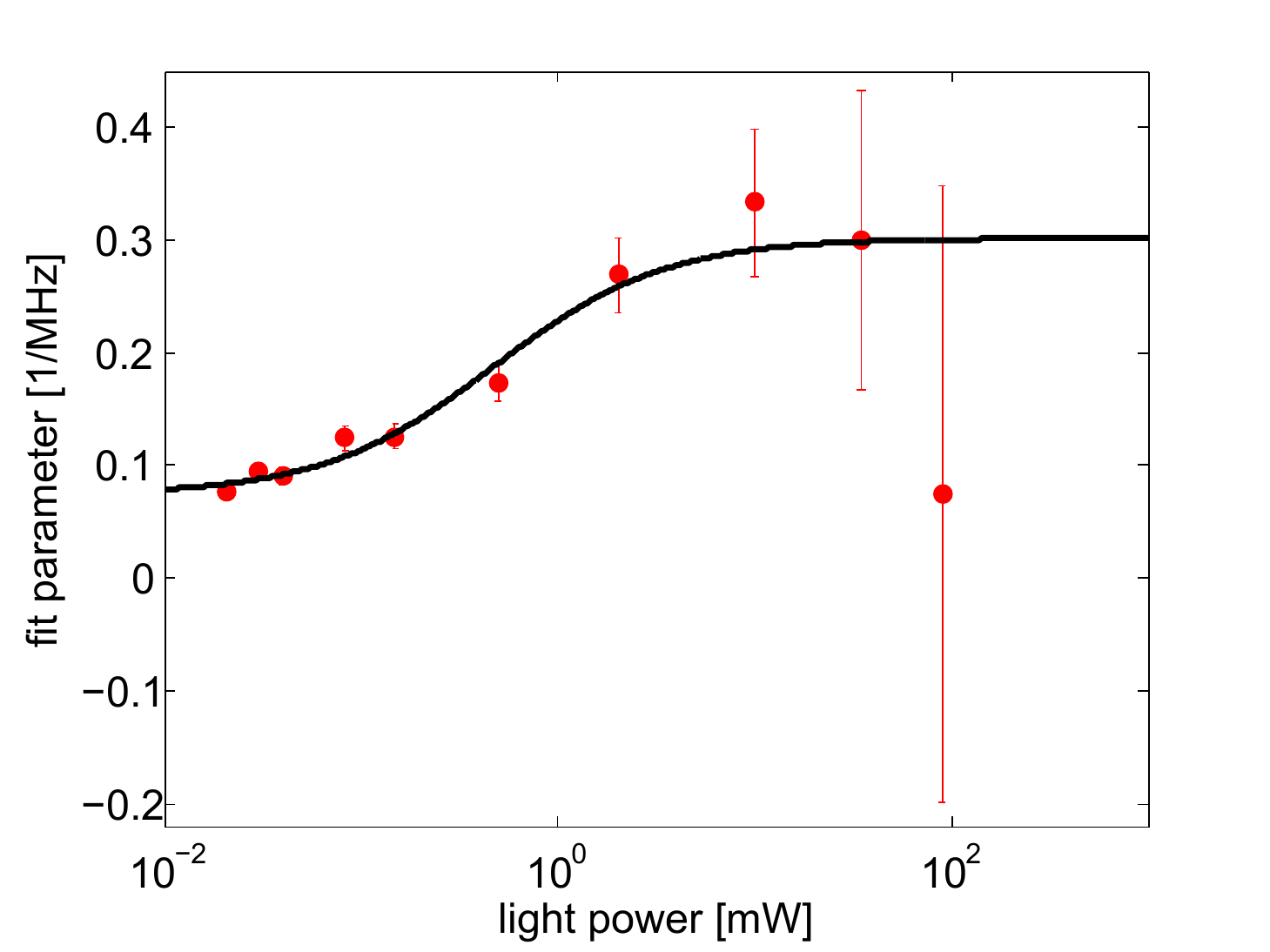}
\caption{Points: Fit parameter $a/\gtwointr$ for different light powers. 
The uncertainties on the points equals the 68\% confidence intervals for the fit parameters $a(P)/\gtwointr$ calculated from the non-linear least squared fit of the measured widths to Eq.\ (\ref{eq:widththeory1}).
Line: Weighted fit of $a(P)/\gtwointr$ to the function 
$ a_1 P/\lr{1+P/b_1}+c_1$ . }
\label{fig:fitparameterC1}
\end{figure}

We choose to make a global fit of the width as a function of Rabi frequency and light power to Eq.\ (\ref{eq:widththeory1}) with the fit parameters: 
$ \left\{ \Delta \nu_{\rm{inh}}, \goneintr/\gtwointr, a/\gtwointr, c/\gtwointr, P_0 \right\}$.
Since data with 12 different light-power settings were fitted, the fit parameter $a(P)/\gtwointr$ will be a vector with 12 components. 
The fits to the experimentally measured widths are shown in Fig.\ \ref{fig:ResultsAndTheory2012}(a) with solid lines and we see that the measured data can be  well fitted by Eq.\ (\ref{eq:widththeory1}).
The values of the fit parameters are given in Table \ref{tab:fitpar} and Fig.\ (\ref{fig:fitparameterC1}). We note that the uncertainty on the fit parameter $a(P)/\gtwointr$ is large for high light powers because the width depends only weakly on $\gamma_{\rm{NV-P1}}$ when $cP \gg \gamma_{\rm{NV-P1}}$.

We emphasize that the term $a\cdot f_R^2/\lr{1+f_R^2/f_0^2}$ in Eq.\ (\ref{eq:widththeory1})
is necessary to obtain good agreement between data and fit since this term is responsible for the negative curvature of the width as a function of Rabi frequency.
The fitted parameter $a/\gtwointr$ is shown in Fig.\ (\ref{fig:fitparameterC1}) and is found to be in the range $\lr{0.1-0.3}$/MHz for the different light-power settings.

The term $a\cdot f_R^2/\lr{1+f_R^2/f_0^2}$ in the fit function for the width 
[Eq.\ (\ref{eq:widththeory1})]
is interpreted as the relaxation rate $\gamma_{\mathrm{NV-P1}}$ due to NV-P1 interaction.
For a Rabi frequency of 1 MHz, fit parameters $a/\gtwointr \approx 0.15/\rm{MHz}$ and $f_0=1.0$ MHz and an estimated $\gtwointr=1/\mu s$, we calculate 
 $\gamma_{\mathrm{NV-P1}} =\gtwointr \cdot \lr{a/\gtwointr} \cdot f_R^2\lr{1+f_R^2/f_0^2}  \approx 0.075\gtwointr \approx 0.075/\rm{\mu s}$.
For a Rabi frequency of 1 MHz, the contribution from $\gamma_{\rm{NV-P1}}$ to the effective $T_1$-relaxation rate $\goneeff$ is therefore much larger than the contribution from the intrinsic relaxation rate which is estimated to be 
$\goneintr \approx 1/\lr{1 \; \mathrm{ms}} =0.001/\mu\mathrm{s}$.
The intrinsic longitudinal spin-relaxation rate is therefore negligible compared to the NV-P1 spin flip rate at high MW powers. 
We conclude that intrinsic $T_1$-relaxation will probably not be important for continuous-wave (CW) magnetometry using NV centers in diamond samples with comparable nitrogen concentration.

We choose to fit the obtained $a(P)/\gtwointr$ to an empirical function 
\begin{equation}
a(P)/\gtwointr=a_1 P/\lr{1+P/b_1}+c_1 .
\label{eq:C1}
\end{equation} 
The fit parameters are given in Table \ref{tab:fitpar} and the fit funtion  plotted in Fig.\ (\ref{fig:fitparameterC1}) as a solid line fits the data well. It is not clear why the parameter $a$ should have this particular dependence on light power.
In order to better understand the process of MW-induced NV-P1 simultaneous spin-flips and the associated spin-flip rate, one would need to further develop a theoretical model and probably do more direct measurements of the NV-P1 spin-flip rate instead of these indirect measurements where the NV-P1 spin flip rate is inferred from the width of the ODMR spectra.

We also note that $\goneintr/\gtwointr$ was fitted to 0.0014(3). This agrees well with the expected value $\approx 10^{-3}$, which was calculated using  estimated intrinsic relaxation rates ($\goneintr\approx 1/\rm{ms}$ and 
$\gtwointr \approx 1/ \rm{\mu s}$). We can also estimate the value of $cP_0$
from the fit parameters. We have 
$cP_0=\gtwointr \cdot\lr{c/\gtwointr}\cdot P_0 \approx \gtwointr \cdot 0.018/\rm{mW}\cdot 39\rm{mW} \approx 0.7\gtwointr\approx 0.7/\rm{\mu s}$. As calculated in Appendix B using a five-level model for ODMR dynamics, $c P_0$ should equal $\Gamma_s$, where $\Gamma_s \approx 1/\lr{200\rm{ns}}=5/ \rm{\mu s}$ is the decay rate from the singlet state. The value of $c P_0$ estimated from the fit parameters is therefore a bit on the low side compared to the value estimated from the singlet state lifetime.


\begin{table}[ht]
\centering
\begin{tabular}{|c|c|}
\hline 
$\Delta \nu_{\rm{inh}}			$     & 3.08(9) MHz \\
$\goneintr/\gtwointr$ 			& 0.0014(3) \\
$a/\gtwointr$         			& see Fig.\ (\ref{fig:fitparameterC1}) \\
$c/\gtwointr$         			& 0.018(5) 1/mW  \\
$P_0$                       & 39(13) mW  \\
$f_0$                       & 1.0(1) MHz  \\
\hline
$a_1$                              & 0.5(2) 1/(MHz $\cdot$ mW) \\
$b_1$                              & 0.5(2) mW \\
$c_1$                              & 0.074(7) 1/MHz \\
\hline
$\theta$                    & $22.9(4)\cdot 10^{-3}$ \\
$\goneintr/c$               & 0.71(6) mW \\
$\gtwointr\cdot \goneintr$  & 0.0047(6) $\lr{\rm{\mu s}}^{-2}$\\
\hline
\end{tabular}
\caption{Fit parameters. 1st column: fit parameters. 2nd column: values for the fit parameters. The fit functions are given by Eqs.\ (\ref{eq:widththeory1}), (\ref{eq:C1}) and (\ref{eq:2levelC2}).}
\label{tab:fitpar}
\end{table}

\subsection{Contrast}
In order to have high magnetic sensitivity, the ODMR resonance should both be narrow and have a high contrast. 
The contrast $\C$ is defined as the total peak height of the ODMR signal: $\C=1-S(\nu_0)/S(\infty)$ [see Eq.\ (\ref{eq:SDelta})], which is larger than the amplitude or peak height $A$ of an individual hyperfine resonance [see Eq.\ (\ref{eq:ODMRsignal})]. We can calculate the relation $\C=A\cdot \left[ 1+ 2g^2/\lr{A_{\rm{hf}}^2+g^2} \right]$ from Eqs.\ (\ref{eq:SDelta}) and (\ref{eq:ODMRsignal}).
When the width is much larger than the hyperfine splitting ($g^2\gg A_{\rm{hf}}^2$) we find a simple relation $\C=3A$.

Figure \ref{fig:ResultsAndTheory2012}(d) shows the fitted amplitudes as a function of light power for several different MW settings. 
The measured amplitudes lie in the range  $\lr{0.1-6.8}\cdot 10^{-3}$. 
The maximum measured contrast is therefore $\C\approx 3\cdot6.8\cdot 10^{-3}\approx2.0\%$. 
As seen in both Fig.\ \ref{fig:ODMR3}(c) and Fig.\ \ref{fig:ResultsAndTheory2012}(c), the contrast increases with MW power. 
The dependence on light power is more complicated [Fig.\ \ref{fig:ResultsAndTheory2012}(d)]. 
First the amplitude increases with light power, then a maximum is reached, and finally the amplitude decreases. Both the maximum amplitude and the light power needed to achieve the maximum amplitude depend on the MW power.

The theoretical expected contrast was calculated using the two-level model described in detail in Appendix A. Upon rearranging, we obtain
\begin{eqnarray}
\C &=& \frac{1}{4} \cdot \theta \cdot 
\frac{cP}{cP+\goneintr \lr{1-\theta}}  \nonumber \\
& \times &
\frac{f_R^2}{f_R^2+ \frac{1}{\lr{2\pi}^2}  \cdot 
\gtwointr \cdot  \left[\goneintr+cP\right]}
\label{eq:2levelC2} .
\end{eqnarray}
Here $\theta$ is an overall proportionality constant defined in Appendix A.
The factor 1/4 is included to take into account that only 1/4 of the NV centers (the ones which are oriented in the direction of the magnetic field) are in resonance with the MW field.  
In Eq.\ (\ref{eq:2levelC2}) for the contrast, the pump rate is assumed to be $\Gamma_P=cP$. 
From Eq.\ (\ref{eq:2levelC2}), we see that the pump rate $\Gamma_P$ should be larger than the $T_1$-relaxation rate $\goneintr$ in order to achieve high contrast [$\Gamma_P > \goneintr \lr{1-\theta}$]. 
If the light power is too low, it will not be possible to polarize the NV centers. This explains the initial increase in contrast with increasing light power.
However, at high light powers 
[$\Gamma_P > \lr{2\pi f_R}^2 / \gtwointr $], 
the optical pumping becomes too fast compared to the MW driving field, and it will not be possible to create a population difference between the two ground states $\ket{0}$ and $\ket{1}$  involved in the MW transition. This explains the decrease in contrast at high light powers.

We choose to make a global fit of the contrast as a function of Rabi frequency and light power to Eq.\ (\ref{eq:2levelC2}) with the fit parameters 
$\left\{\theta, \goneintr/c, \goneintr \cdot \gtwointr \right\}$.
The fit to the data are shown in 
Fig.\ \ref{fig:ResultsAndTheory2012} (c), (d) as solid lines and the fit parameters are given in  Table \ref{tab:fitpar}. 
The fit function Eq.\ (\ref{eq:2levelC2}) agrees reasonably well with the data.
The discrepancy between data and fit is probably due to the simplicity of the two-level model leading to Eq.\ (\ref{eq:2levelC2}) for the contrast, where we did not include the effect of inhomogeneous broadening, MW-induced simultaneous spin-flips of NV and P1 centers, and possible population of the singlet and excited states.

\subsection{Magnetometer sensitivity}
Using a measured ODMR spectrum,
one can calculate the experimental magnetometer sensitivity 
from the uncertainty on the fitted center frequency and the averaging time used to obtain the ODMR spectrum.
However, since our measurement uncertainty is mainly limited by technical noise specific to our apparatus, this will not tell us about the fundamental limits to the sensitivity.
We instead investigate the projected sensitivity for a shot-noise limited measurement.

The shot-noise limited sensitivity is of the order of
\begin{equation}
S_B \approx   \frac{2\pi}{\gamma} \cdot \frac{\Delta \nu}{\C\sqrt{\mathcal{R}}},
\label{eq:SB}
\end{equation}
where $\Delta \nu$ is the FWHM of the magnetic resonance, $\gamma=1.761\cdot10^{11} s^{-1}T^{-1}$ is the gyromagnetic ratio for the NV center 
and $\mathcal{R}$ is the rate of detected photons.
The experimental sensitivity $S_B$ is calculated using Eq.\ (\ref{eq:SB}) and experimentally measured values of $\Delta \nu$, $\C$ 
and $\mathcal{R}$. 
When calculating $S_B$, $\Delta \nu$ was set to the fitted FWHM, $\C$ was set to $3A$ (these assumptions are valid when the hyperfine structure is not resolved)  
and $\mathcal{R}$ was estimated from the measured power of the fluorescence.
We can calculate $\mathcal{R}=P_{fl}/E_{fl}$ where $E_{fl}=hc/\lambda$ ($h$ Planck's constant, $c$ speed of light, $\lambda \approx 670$ nm wavelength of the fluorescence). Using the calibration shown in Fig.\ \ref{fig:setuplevels}(d), we find that 1 mW of green light produces 6.2 $\mu$W of red fluorescence which corresponds to a rate of detected photons $\mathcal{R}=2.1\cdot 10^{13}\;\mathrm{s}^{-1}$.

\begin{figure}
	\centering
\includegraphics[width=.44\textwidth]{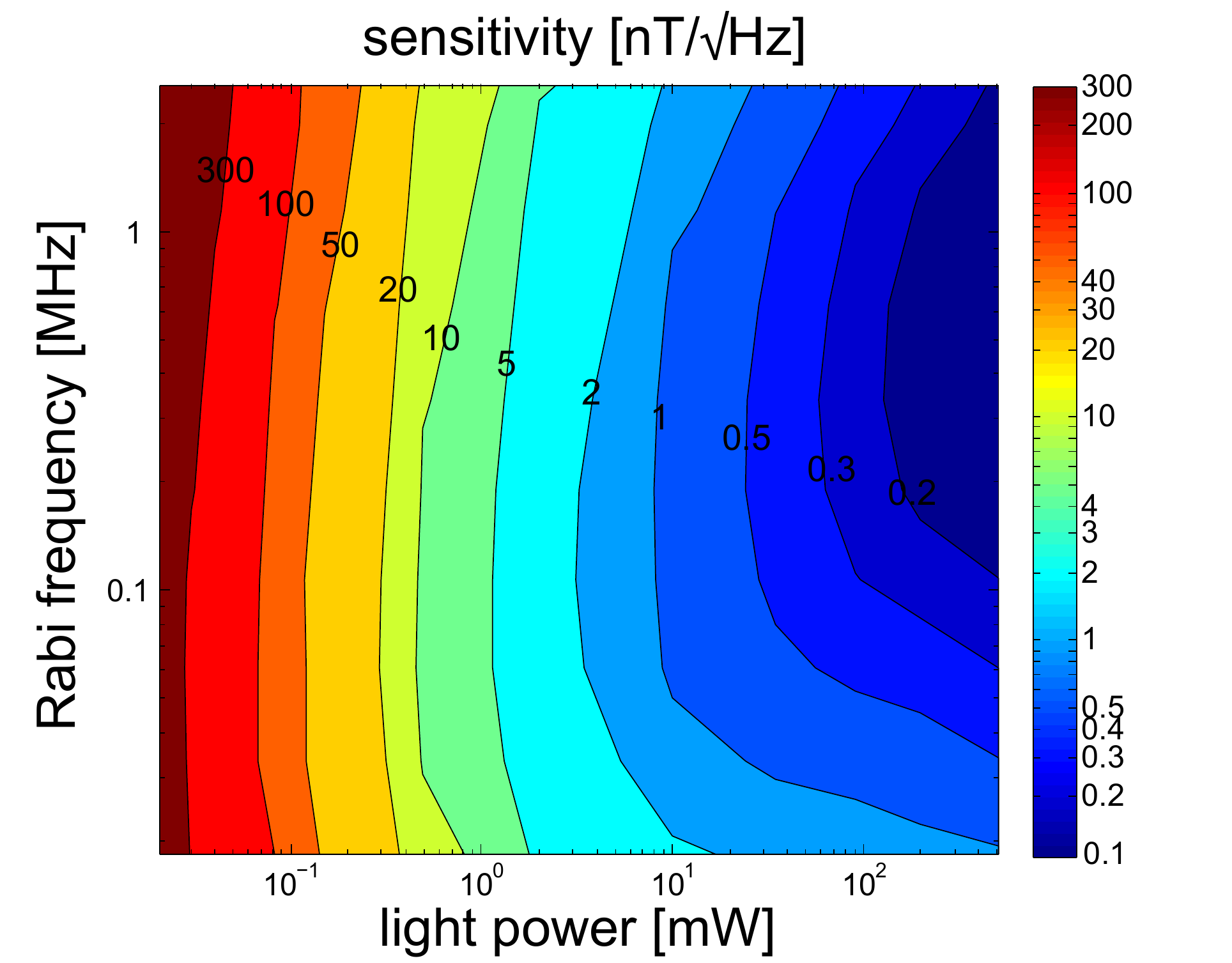}
\caption{
Contour plot of the experimental magnetic field-sensitivity $S_B$  as a function of light power $P$ and Rabi frequency $f_R$.
Notice the logarithmic scales.
}
\label{fig:sensitivity}
\end{figure}

Figure \ref{fig:sensitivity} shows a contour plot of $S_B$ as a function of $P$ and $f_R$. 
In total, experimental data with more than 100 different settings of $(P,f_R)$ were used to create the contour plot.
The best sensitivity
$S_B\approx 0.1$ nT/$\sqrt{\rm{Hz}}$ is reached for the highest light power $P=500$ mW and Rabi frequency settings  $f_R= 0.34$, 0.62 and 1.14 MHz. 
For even higher light powers ($P>500$ mW), the sensitivity is expected to eventually become worse \cite{Dreau11}.
We note that our best projected shot-noise limited sensitivity is significantly better than the actual measured sensitivity obtained with ensembles of NV centers so far \cite{Chang12,Acosta10APL}.
%


When detecting fluorescence from a diamond with a confocal microscope, the photon detection efficiency is rather small (in our case $\approx 0.8$ \%) due to the high index of refraction of diamond ($n \approx 2.4$).
We note that there exist several methods to increase the detection efficiency including the ``side-collection'' technique \cite{Sage12}, the use of solid immersion lenses \cite{Hadden10,Siyushev10} and a detection scheme based on IR absorption \cite{Acosta10APL}. 
If such techniques would be employed in our experiment, the detection efficiency and thereby the rate of detected photons could be increased by roughly two orders of magnitude. 
This would lead to an overall improvement of our best projected shot-noise limited sensitivity by a factor of 10 to the value 
10 $ \mathrm{pT}/ \sqrt{\mathrm{Hz}}$.

Combining the fact that the settings of $(P,f_R)$ where the sensitivity is optimal [Fig.\ \ref{fig:sensitivity}], are also the settings where the width of the magnetic resonance is reduced by the light-narrowing effect 
[Fig.\ \ref{fig:ResultsAndTheory2012}(b)], we reach the conclusion that the light-narrowing effect plays an important role for optimal sensitive  magnetometry with NV centers in diamond.

\subsection{Light narrowing and IR absorption}
We have so far focused on ODMR signals detected using fluorescence from the spin-triplet excited states. An alternative way of measuring magnetic fields using NV centers in diamond is to use infrared (IR) absorption at the singlet transition at 1042 nm \cite{Acosta10PRB,Acosta10APL}. 
In this case, a green laser is used to pump NV centers into the lower spin-singlet state, the MW frequency is scanned around magnetic resonance while the absorption of an IR laser (1042 nm) is measured. 

We have analyzed a five-level model for the NV center. The details of the model are given in Appendix B. As shown in Fig.\ \ref{fig:fivelevels}, the model includes two ground states $\ket{0},\ket{1}$, two excited states $\ket{e0},\ket{e1}$ and one state $\ket{s}$ representing the singlet states. 
The IR absorption is proportional to the number of NV centers in the singlet state $a_{\mathrm{IR}} \propto \rho_{ss}$ (here $\rho_{ss}$ denotes the diagonal density-matrix element associated with the state $\ket{s}$). Using Bloch equations (see Appendix B), one can calculate 
that the steady-state solution $\rho_{ss}(\Delta)$, where $\Delta$ is the MW detuning, will have the same lineshape and the same width as the ODMR signal $S(\Delta)$ measured using fluorescence. 
Light narrowing should therefore also occur when measuring IR absorption (in the sense that the width of the ODMR signal measured using IR absorption decreases with increasing green pump power). 
This makes sense since light narrowing of the  MW resonance should not depend on the detection method (fluoresence vs absorption).

\section{Conclusions}
In this paper, the light-narrowing effect was discussed using a two-level model, which is sufficient to demonstrate the basic phenomenon. 
We experimentally demonstrated light narrowing using NV centers in diamond and found a reduction in the linewidth of the microwave transition by more than a factor of two when the light power was increased. 
The obtained reduction in linewidth is comparable to what was found in earlier studies of light narrowing \cite{Bhaskar81,Appelt99} in alkali vapor cells.
It was also found that MW-induced simultaneous spin-flips of NV centers and P1 centers are important for determining the lineshape and width of the ODMR signal.
Finally, we showed that light narrowing is an essential factor in optimizing the sensitivity of magnetometers based on NV centers in diamond.

\begin{acknowledgments}
This work was supported by AFOSR/DARPA, IMOD, NSF, and the NATO Science for Peace program. 
K. J. was supported by the Danish Council for Independent Research $|$ Natural Sciences.
The authors acknowledge Erik Bauch for initial contributions to this work.
\end{acknowledgments}
\appendix

\section{Two-level model}

We analyze the situation of a two-level atom driven by a MW field [see Fig.\ \ref{fig:setuplevels}(a)] with detuning $\Delta$ and power corresponding to the on-resonance Rabi frequency $\Omega_R \;$ \cite{Lasers}. $\Delta$ and $\Omega_R$ are defined to be angular frequencies measured in rad/s.
We assume that the upper level can be optically pumped into the lower level with a rate $\Gp$.
The populations of the $\ket{0}$ and $\ket{1}$ states are assumed to decay towards equal populations at a rate $\gamma_1 = 1/T_1$.
The transverse relaxation rate is denoted $\gamma_2 = 1/T_2$. 
We have the following equations of motion (the Bloch equations) for the density-matrix elements:
\begin{subequations}
\small
\begin{align}
\dot{\rho}_{00} =& -\frac{i \Omega_R }{2} \lr{\rho_{01}- \rho_{10}} 
-\frac{\gamma_1}{2} \lr{ \rho_{00}-\rho_{11} } +\Gp \rho_{11} 
\label{eq:Rabi2levelGp} \\
\dot{\rho}_{11} =& +\frac{i \Omega_R }{2} \lr{ \rho_{01}- \rho_{10}} 
-\frac{\gamma_1}{2} \lr{ \rho_{11}-\rho_{00} } - \Gp \rho_{11}
\label{eq:Rabi2levelBGp} \\
\dot{\rho}_{01} =& -\lr{\gtwoeff-i\Delta} \rho_{01}+i\frac{\Omega_R}{2}\lr{\rho_{11}-\rho_{00}} 
\label{eq:Rabi2levelCGp}   \\
\dot{\rho}_{10} =& -\lr{\gtwoeff+i\Delta}  \rho_{10}-i\frac{\Omega_R}{2}\lr{\rho_{11}-\rho_{00}}.
\label{eq:Rabi2levelDGp}
\end{align}
\normalsize
\end{subequations}
The effective dephasing rate is denoted $\gtwoeff = \gamma_2+\Gp/2$.
The steady-state solutions can be calculated from Eqs.\ (\ref{eq:Rabi2levelGp})--(\ref{eq:Rabi2levelDGp}).
Similar to what was done in Ref.\ \onlinecite{Dreau11}, we assume that the ODMR signal can be written as $S(\Delta)=\alpha\rho_{00}+\beta \rho_{11}$, where $\alpha>\beta$. In this case, the ODMR signal will have the following form:
\begin{equation}
S(\Delta)=S(\infty)\left[ 1-\frac{\C\gamma^2}{\Delta^2+\gamma^2} \right].
\label{eq:2levellineshape}
\end{equation}
$S(\infty)$ is the signal when the MW's are off-resonant 
($\abs{\Delta} \rightarrow \infty$) and 
$\gamma$ is the hwhm in rad/s.
We calculate the FWHM in Hz: $\Delta \nu=\gamma/\pi$ to be
\begin{equation}
\Delta \nu=\sqrt{\lr{\frac{\gtwoeff}{\pi}}^2+\frac{4\gtwoeff}{\gamma_1+\Gp} \cdot f_R^2},
\label{eq:twolevelwidth}
\end{equation}
where $f_R=\Omega_R/\lr{2\pi}$ is the Rabi frequency in Hz.
The ODMR signal contrast is
\begin{equation}
\C=\theta \cdot \frac{\Gp}{\Gp+\gamma_1 \lr{1-\theta}} \cdot
\frac{\Omega_R^2}{\Omega_R^2+\gtwoeff \lr{\gamma_1+\Gp}},
\label{eq:2levelC}
\end{equation}
where $\theta=\lr{\alpha-\beta}/\lr{2\alpha}$.
Note that similar expressions for the linewidth and contrast have also been derived in the Appendix of Ref.\ \onlinecite{Dreau11}.
\section{Five-level model}

\begin{figure}
	\centering
\includegraphics[width=.4\textwidth]{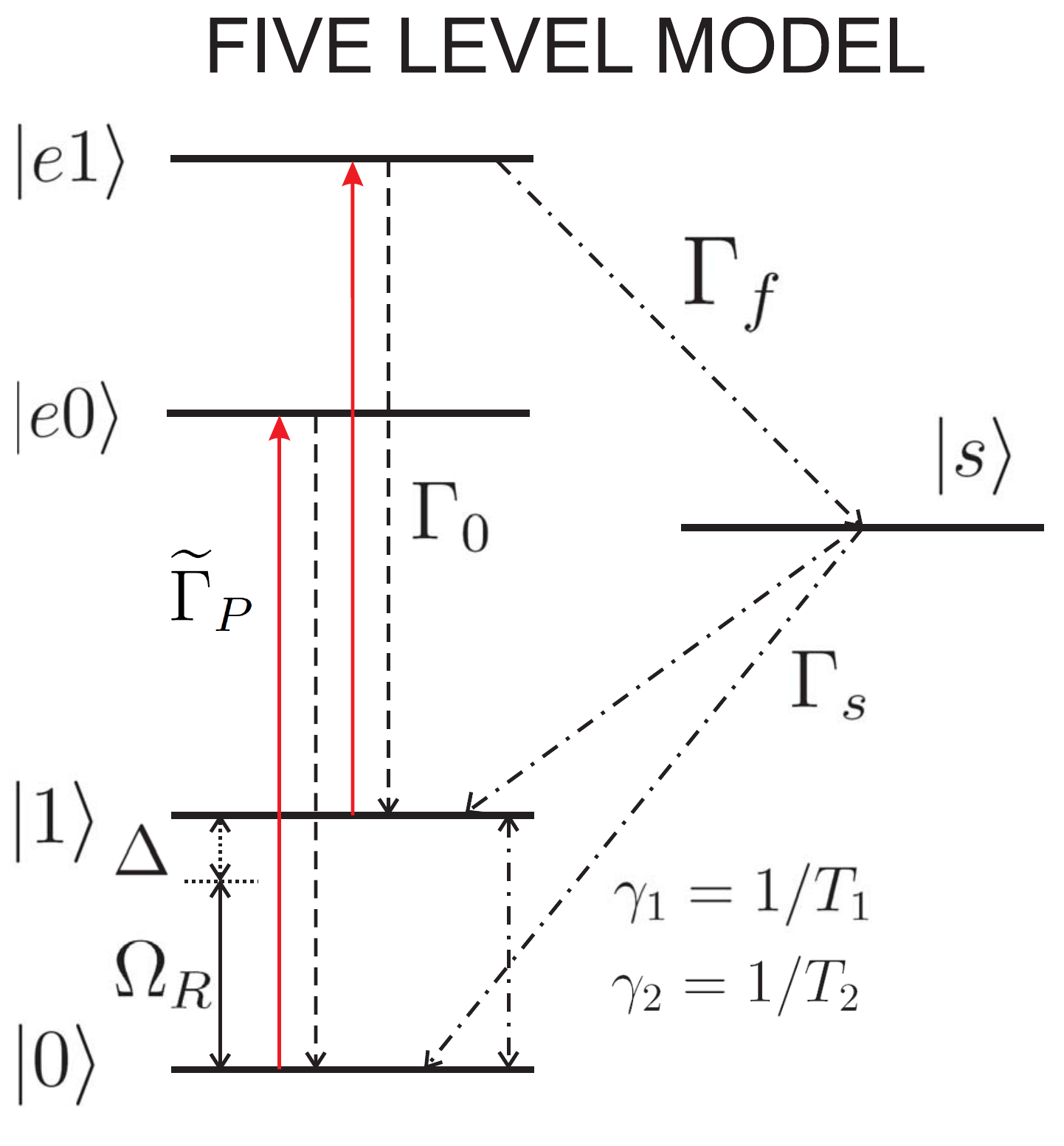}
\caption{Five-level model of ODMR dynamics.
The two ground states $\ket{0}$ and $\ket{1}$ are coupled with a MW field (Rabi frequency $\Omega_R$ and detuning $\Delta$). 
Laser light excite the states $\ket{0}$ and $\ket{1}$ with equal rate $\Gptilde$ to the excited states $\ket{e0}$ and $\ket{e1}$, respectively.
The excited states decay to the ground state with rate $\Gamma_0$. The state $\ket{e1}$ can also decay to the singlet state $\ket{s}$ with rate $\Gamma_f$. The singlet state $\ket{s}$ decays with a rate $\Gamma_s$ to either  $\ket{0}$ or $\ket{1}$.}
\label{fig:fivelevels}
\end{figure}

We now present a five-level model describing light narrowing and optical pumping of the NV center. The model includes the ground states, excited states and singlet states of the NV center 
(see Fig.\ \ref{fig:fivelevels}), and is therefore more detailed than the two-level model [see Fig.\ \ref{fig:setuplevels}(a)] described earlier.
Previously, a five level model has been used for calculating the dynamics of a single NV center \cite{Robledo11}. We will use the five-level model to calculate the lineshape and width of ODMR spectra based on both fluoresence and infrared absorption detection.

As in the two-level model, two ground states $\ket{0}$ and $\ket{1}$  are coupled by a MW field with Rabi frequency $\Omega_R$ and detuning $\Delta$.
Laser light can excite the ground states with a rate $\Gptilde$ through a spin-conserving transition to the excited states which in the model are denoted $\ket{e0}$ and $\ket{e1}$. 
The excited states can decay with rates $\Gamma_0$ to the ground states. In the model, the $\ket{e1}$ state can also decay with rate $\Gamma_f$ to the state $\ket{s}$ representing the singlet states. Finally, the state $\ket{s}$ can decay with a rate $\Gamma_s$ to the ground states $\ket{0}$ and $\ket{1}$ with equal probability.
Realistic values of the decay rates are $\Gamma_0=1/(12 \; \rm{ns})$, $\Gamma_f=1/(12 \;\rm{ns})$ and $\Gamma_s=1/(200\; \rm{ns})$.

The equations of motion for the density-matrix elements for the five-level system $\left\{\ket{0},\ket{1},\ket{e0},\ket{e1},\ket{s} \right\}$ can be written as:
\begin{subequations}
\begin{align}
& \dot{\rho}_{00} &=& \quad -\frac{i \Omega_R }{2} \lr{\rho_{01}- \rho_{10}} 
- \frac{\gamma_1}{2} \lr{\rho_{00} -\rho_{11} }
\nonumber \\
& & & \quad -\Gptilde \rho_{00} +\Gamma_0 \rho_{e0e0}+ \frac{\Gamma_s}{2} \rho_{ss} \label{eq:5levelmodel} \\
& \dot{\rho}_{11}   &=& \quad +\frac{i \Omega_R }{2} \lr{\rho_{01}- \rho_{10}} 
-\frac{\gamma_1}{2} \lr{\rho_{11} -\rho_{00} } \nonumber \\
& & & \quad -\Gptilde \rho_{11} +\Gamma_0 \rho_{e1e1} + \frac{\Gamma_s}{2} \rho_{ss}
 \label{eq:5levelmodelB} \\
& \dot{\rho}_{01} &=& \quad -\lr{\gtwoeff-i\Delta} \rho_{01} 
 +i\frac{\Omega_R}{2}\lr{\rho_{11}-\rho_{00}} 
\label{eq:5levelmodelC} \\
& \dot{\rho}_{10}  &=& \quad -\lr{\gtwoeff+i\Delta} \rho_{10}
-i\frac{\Omega_R}{2}\lr{\rho_{11}-\rho_{00}} 
\label{eq:5levelmodelD} \\
& \dot{\rho}_{e0e0} &=& \quad +\Gptilde \rho_{00} -\Gamma_0 \rho_{e0e0} 
\label{eq:5levelmodelE} \\
& \dot{\rho}_{e1e1} &=& \quad +\Gptilde \rho_{11} -\Gamma_0 \rho_{e1e1} -\Gamma_f \rho_{e1e1} 
\label{eq:5levelmodelF} \\
& \dot{\rho}_{ss}  &=& \quad +\Gamma_f \rho_{e1e1} - \Gamma_s \rho_{ss}.
\label{eq:5levelmodelG}
\end{align}
\end{subequations}
The $T_1$-decay is here assumed to be towards equal population in $\ket{0}$ and $\ket{1}$. The transverse relaxation rate is denoted $\gamma_2=1/T_2$, and the effective transverse relaxation rate is defined as $\gtwoeff=\gamma_2+\Gptilde/2$.
The NV centers are pumped on the spin-conserving transitions $\ket{0}\rightarrow \ket{e0}$ and $\ket{1}\rightarrow \ket{e1}$ with off-resonant green 532 nm light.
Coherences between ground and excited states are set to zero since we use off-resonant excitation which does not preserve such coherences.

The steady-state solutions for the above density-matrix elements can be calculated analytically from Eqs.\ (\ref{eq:5levelmodel})--(\ref{eq:5levelmodelG}).
If one measures the amount of fluorescence from the spin-triplet excited states, the ODMR signal will be on the form
\begin{equation}
S(\Delta) \propto \rho_{e0e0}+\frac{\Gamma_0}{\Gamma_0+\Gamma_f} \rho_{e1e1}
=
S(\infty)\left[1- \frac{\C \gamma^2}{\Delta^2+\gamma^2} \right],
\label{eq:5levellineshape}
\end{equation}
where $\C$ is the contrast, $\gamma$ the hwhm in rad/s, and $S(\infty)$ is the obtained signal when the MW's are far off resonance.
The lineshape given by Eq.\ (\ref{eq:5levellineshape}) is the same as was obtained with the two-level model [Eq.\ (\ref{eq:2levellineshape})].
We calculate the FWHM in Hz
which under the assumptions 
$\Gamma_0\approx \Gamma_f$,  
$\Gptilde\ll \Gamma_0$
and
$\gamma_1 \ll \Gamma_s$
has the simple form
\begin{equation} 
\Delta \nu = 
\sqrt{
\lr{ \frac{\gtwoeff}{\pi}   }^2 +
\frac{ 4 \gtwoeff \cdot \left[1+\Gptilde/\lr{4\Gamma_s} \right] }{\gamma_1+\Gptilde/4} \cdot f_R^2 
}.
\label{eq:g5level}
\end{equation}
The assumption $\Gptilde\ll \Gamma_0$ means that the light power should be sufficiently low such that the excited states are not significantly populated. This assumption should be valid in our experiments, since we do not observe much saturation of the amount of fluorescence with increasing light power [see Fig.\ \ref{fig:setuplevels}(d)].

For low excitation power: $\Gptilde/4 \ll \Gamma_s$, the two-level and the five-level models give identical results for the width [Eqs.\ (\ref{eq:twolevelwidth}) and (\ref{eq:g5level})], if one associates the five-level excitation rate diveded by four with the two-level optical pumping rate: $\Gptilde/4 \sim \Gamma_P$. 
For higher excitation powers: $\Gptilde/4 \geq \Gamma_s $, the singlet states become partially occupied, leading to saturation effects. 
This is included in the five-level formula for the width 
[Eq.\ (\ref{eq:g5level})] by the term $\left[1+\Gptilde/\lr{4\Gamma_s} \right]$.

Equation (\ref{eq:g5level}) can be re-written in terms of light power as 
\begin{equation} 
\Delta \nu = 
\sqrt{
\lr{ \frac{\gtwoeff}{\pi}   }^2 +
\frac{ 4 \gtwoeff \cdot \left[1+P/P_0 \right] }{\gamma_1+cP} \cdot f_R^2 
}.
\label{eq:g5levelPower}
\end{equation}
In the above we use that the excitation rate is proportional to light power $\Gptilde = 4cP$. The proportionality constant is denoted $4c$. The optical pumping saturation power $P_0$ due to population of the singlet states is defined by the equation $c P_0 = \Gamma_s$.

Instead of detecting fluorescence emitted from the spin-triplet excited states, one can measure ODMR signals using absorption at the infrared (IR) transition at 1042 nm involving the spin-singlet states \cite{Acosta10PRB,Acosta10APL}. In this case, a green laser is used to pump NV centers into the lower spin-singlet state, the MW frequency is scanned around magnetic resonance while the absorption of an IR laser (1042 nm) is measured. The IR absorption will be proportional to the number of NV centers in the singlet state $a_{\mathrm{IR}} \propto \rho_{ss}$. One can calculate from 
Eqs.\ (\ref{eq:5levelmodel})--(\ref{eq:5levelmodelG})
that the steady-state solution $\rho_{ss}(\Delta)$ will have the same lineshape [Eq.\ (\ref{eq:5levellineshape})] and the same width [Eq.\ (\ref{eq:g5level})] as the ODMR signal $S(\Delta)$ measured using fluorescence. 
Light narrowing can therefore also occur when measuring IR absorption (in the sense that the width of the ODMR signal measured using IR absorption decrease with increasing green pump power).


\end{document}